# Optical phase conjugation with less than a photon per degree of freedom

M. Jang[1,†], C. Yang[1,*], I.M. Vellekoop[2]

[1]*Electrical Engineering, California Institute of Technology, 1200 East California Boulevard, Pasadena, California, 91125, USA*

[2]*Biomedical Photonic Imaging Group, MIRA Institute for Biomedical Technology & Technical Medicine, University of Twente, PO Box 217, 7500 AE Enschede, The Netherlands*

## Abstract

We demonstrate experimentally that optical phase conjugation can be used to focus light through strongly scattering media even when far less than a photon per optical degree of freedom is detected. We found that the best achievable intensity contrast is equal to the total number of detected photons, as long as the resolution of the system is high enough. Our results demonstrate that phase conjugation can be used even when the photon budget is extremely low, such as in high-speed focusing through dynamic media, or imaging deep inside tissue.

## MAIN TEXT

Biological tissue strongly scatters light, traditionally limiting the depth of optical imaging to within one millimeter. Recently, methods where phase conjugate fields "time-reverse" the effect of scattering and converge on a guide star, have been shown effective in overcoming scattering [1–12], thus opening an avenue towards high-resolution optical imaging and manipulation in biological issue. A general concern, however, is that the guide stars may be very weak, leading to the question of how many photons need to be detected to effectively phase conjugate a light wave. Here we demonstrate that phase conjugation is effective even when less than a photon per optical degree of freedom is detected. We found that the best achievable intensity contrast is equal to the total number of detected photons, as long as the resolution of the system is high enough.

Modern optical phase conjugation systems are commonly based on digital optical phase conjugation (DOPC) [7,8,13,14] or use 'analog' photorefractive materials [15–18]. Both approaches are two-step processes. In the 'recording' step, light propagates from a guide star to the phase conjugation system. In DOPC, the phase and amplitude of the scattered wave ($E^+$) is typically measured using phase stepping or off-axis holography using a reference beam. Likewise, in the analog approach, light interferes with a reference beam in order to generate a hologram in the photorefractive material. In the 'playback' step, the system generates the phase conjugate copy of the complex field $E^- \propto (E^+)^*$, either by using a digital spatial light modulator (SLM), or by illuminating the photorefractive material with a readout beam. Due to the time-reversal symmetry of light propagation, the phase-conjugated wave propagates back through the scattering medium and focuses at the guide star.

In the standard view, phase conjugation increases the intensity in the desired focus by an average factor of $\eta = M$, where $M$ is the number of independently controlled optical modes [1,13,19]. This result assumes that the fields are measured and reproduced exactly. However, until now it remained unclear how phase conjugation performs in the low photon limit, where shot noise prevents an accurate measurement of $E^+$ [20]. One may think that the number of detected photons for each optical mode needs to be high enough to measure the field in each of the $M$ modes with sufficient accuracy. However, here we demonstrate the opposite: phase conjugation is still possible even with far less than a single detected photon per degree of freedom. Our study reveals that the fundamental limit for the best possible enhancement $\eta$ is given by

$$\bar{\eta}_{max} = \frac{1}{M^{-1} + \bar{n}_r^{-1} + \bar{n}_s^{-1}} \tag{1}$$

where $\bar{n}_r$ and $\bar{n}_s$ are the total number of detected photons coming from the reference and scattered beam, respectively, and the overline denotes averaging over repetitions of the measurement. Interestingly, in the low photon limit (i.e., when $\bar{n}_s \ll M$), the enhancement is equal to the total number of detected signal photons (i.e. $\bar{\eta}_{max} = \bar{n}_s$).

In order to derive Eq. (1), we first introduce the fidelity $|\gamma|^2$ [21], where

$$\gamma = \frac{\sum_{m=1}^{M} E_m^- E_m^+}{\sqrt{\sum_{m=1}^{M} |E_m^-|^2} \sqrt{\sum_{m=1}^{M} |E_m^+|^2}} \tag{2}$$

with $m$ the index of the optical mode. The fidelity denotes the fraction of the incident power that is shaped correctly and, hence, contributes to the focus. The enhancement and fidelity are related through $\eta = |\gamma|^2 M$ [21]. Ideally, $E^- \propto (E^+)^*$, and $|\gamma|^2 = 1$.

In practice, however, shot noise limits the accuracy at which $E^+$ can be measured. We model the number of detected photons in a single measurement as $n = \bar{n} + \xi$, where $\bar{n}$ is the average photon count, and $\xi$ is a noise term with the statistical properties $\bar{\xi} = 0$ and $\overline{\xi^2} = \bar{n}$ following from Poisson statistics [22].

In a phase-stepping interferometry setup, the number of detected photons in a single optical mode is the result of interference between the scattered light and the reference beam

$$n_{k,m} = \frac{\bar{n}_{r,m}}{K} + \frac{\bar{n}_{s,m}}{K} + \frac{2}{K}\sqrt{\bar{n}_{r,m}\bar{n}_{s,m}}\cos(\phi_{0,m} + 2\pi k/K) + \xi_{k,m} = \bar{n}_{k,m} + \xi_{k,m}. \quad (3)$$

$\phi_{0,m}$ is the relative phase difference between two beams at a chosen mode $m$. $K$ is the total number of phase steps and $k \in 1..K$, denotes the step number. For off-axis holography, the relation is identical: to measure the amplitude and phase of a single optical mode, $K$ different pixels are needed; when we identify these pixels by $k \in 1..K$, the same equation is found. The reference intensity is considered to be homogeneous across all optical modes: $\bar{n}_{r,m} = \bar{n}_r/M$. The complex amplitude of the signal beam is then reconstructed using

$$E_m^- = \sum_{k=1}^{K} n_{k,m} e^{-i2\pi k/K} \quad (4)$$

By substituting Eq. (3) and averaging over measurements, we find

$$\bar{E}_m^- = \sqrt{\bar{n}_{r,m}\bar{n}_{s,m}}\, e^{i\phi_{0,m}} = \sqrt{\bar{n}_{r,m}}\, E_m^{+*} \quad (5)$$

and

$$\overline{|E_m^-|^2} = \bar{n}_{r,m}\bar{n}_{s,m} + \sum_{k=1}^{K} \overline{\xi_{k,m}^2} = \bar{n}_{r,m}\bar{n}_{s,m} + \bar{n}_{r,m} + \bar{n}_{s,m} \quad (6)$$

Assuming that $M$ is sufficiently high, we can write $\sum_{m=1}^{M} E_m^- E_m^+ \approx \sum_{m=1}^{M} \bar{E}_m^- E_m^+$ and $\sum_{m=1}^{M} |E_m^-|^2 \approx \sum_{m=1}^{M} \overline{|E_m^-|^2}$. We can now substitute Eq. (5) and (6) into Eq. (2) to find

$$|\gamma|^2 = \frac{\left(\sum_{m=1}^{M} \bar{n}_{s,m}\right)^2 \bar{n}_r / M}{\left(\sum_{m=1}^{M} \bar{n}_{r,m} \bar{n}_{s,m} + \bar{n}_{r,m} + \bar{n}_{s,m}\right)\left(\sum_{m=1}^{M} \bar{n}_{s,m}\right)} = \frac{\bar{n}_s^2 \bar{n}_r / M}{\left(\bar{n}_s \bar{n}_r / M + \bar{n}_r + \bar{n}_s\right) \bar{n}_s} = \frac{M^{-1}}{M^{-1} + \bar{n}_r^{-1} + \bar{n}_s^{-1}} \qquad (7)$$

which gives us the result in Eq. (1). Interestingly, in the low photon limit, the fidelity equals $\bar{n}_s / M$, so increasing the number of degrees of freedom reduces the fidelity, while the enhancement $\eta = |\gamma|^2 M$ in the focus remains constant.

## Experiment

In our experiment, digital optical phase conjugation (DOPC) [7,8,13,14] is used to generate the phase conjugate beam (see Fig. 1 for a detailed schematic of the experiment). The digital implementation here is particularly useful to quantify the scattered input photon number $\bar{n}_s$ and to see how it affects the phase conjugation results.

Our experiment is designed to observe the contrast $\eta$ over a wide range of $\bar{n}_s$ (see Fig. 1). A 532-nm laser beam is split into three beam paths: input beam, reference beam, and playback beam. All beams are spatially filtered using single mode fibers. Beam splitter BS2 divides the DOPC system into a detection part (CMOS and BS1) and a playback part (SLM and BS3). The SLM is placed in the exact mirror image of the image sensor.

In the recording step, the collimated input beam, scattered by a 450 μm-thick opaque opal glass (#43-717, Edmund Optics), was interfered with a plane reference beam on the image sensor (PCO.edge 5.5, PCO; 16-bit sCMOS sensor; readout noise 1.5e- (rms), dark noise 0.5e-/s per pixel; quantum efficiency ~50%). The number of controlled modes $M$ was around 220,000, with $\bar{n}_r / M \sim 10,000$ reference photons per optical mode. The average number of signal photons per optical mode $\bar{n}_s / M$, which was controlled with a pre-calibrated acousto-optic modulator (not shown in Fig. 1), was within the range of 0.004-2000. The input wavefront is measured using phase-stepping holography with the four phase steps 0, $\pi/2$, $\pi$, and $3\pi/2$.

In the playback step, the phase conjugate copy of the measured wavefront was generated using a phase-only spatial light modulator (PLUTO phase only SLM, Holoeye). Then, the fraction of power in the phase conjugate beam, coupled back to the single mode fiber through the opal glass, was monitored by the APD. The use of a single mode fiber assures that the optical mode we monitor is in the exact time-reversal symmetry with the input mode. The intensity distribution of the phase conjugate beam was also directly visualized by the CCD sensor.

Even if the incident field could be measured exactly, the accuracy of the reproduction is limited by experimental factors, such as the fact that a phase-only SLM is used instead of the idealized phase-and-amplitude light modulator. Other factors, such as pixel cross-talk on the SLM, or imperfect alignment further reduce the fidelity. The effect of these experimental factors and the effect of shot noise will be statistically independent. Therefore, we can define an experimental fidelity as $|\gamma_{ex}|^2 \equiv \alpha_{ex}|\gamma|^2$, where $|\gamma|^2$ is given by Eq. (7), and the constant $\alpha_{ex}$ incorporates all experimental limitations (as discussed in Ref. [23]). For perfect phase-only modulation $\alpha_{ex} = \pi/4$ [19], and typical experimental values range from 0.1-0.5 [13,14]. The experimental constant $\alpha_{ex}$ in our experiment was determined to be 0.27, using a high-power input beam so that the effects of shot noise are negligible.

The experiment confirms our theoretical predictions. In Fig. 2, we plotted the experimental enhancement versus $\bar{n}_s$ together with the theoretical curve $\eta \equiv \alpha_{ex}|\gamma|^2 M$. The experimental enhancement scales with the total number of signal photons $\bar{n}_s$ exactly as expected over a scaling range of over five orders of magnitude. The image captured from the CCD sensor further confirms that the time-reversed mode is reconstructed with a significant contrast corresponding to the scaling relation even with the extremely low value of 0.004 photon per mode, corresponding to about 880 signal photons in total.

In order to have a better understanding of the robustness in the low photon limit, we further analyzed the influence of shot noise on the phase measurement accuracy for the case that $K = 4$. By letting $E_m^- = A_m \cos\phi_m + iA_m \sin\phi_m$, the joint probability density function (PDF) for the measured amplitude and phase is given by

$$P_{A,\phi}\left(A_m, \phi_m \middle| \bar{n}_{1,m}, \bar{n}_{2,m}, \bar{n}_{3,m}, \bar{n}_{4,m}\right) = A_m P_{\Delta}\left(A_m \cos\phi_m \middle| \bar{n}_{4,m}, \bar{n}_{2,m}\right) P_{\Delta}\left(A_m \sin\phi_m \middle| \bar{n}_{1,m}, \bar{n}_{3,m}\right) \quad (8)$$

where $P_{\Delta}\left(\Delta \middle| \bar{n}_1, \bar{n}_2\right)$ is the probability mass function of the subtraction of two independent Poissonian random variables with means $\bar{n}_1$ and $\bar{n}_2$ respectively, which is known as the Skellam distribution [24]

$$P_{\Delta}\left(\Delta \middle| \bar{n}_1, \bar{n}_2\right) = e^{-(\bar{n}_1+\bar{n}_2)}\left(\frac{\bar{n}_1}{\bar{n}_2}\right)^{\Delta/2} I_{\Delta}\left(2\sqrt{\bar{n}_1 \bar{n}_2}\right) \quad (9)$$

where $I_\Delta$ is the modified Bessel function of the first kind. To calculate the distribution of the measured phase, we choose an input beam with $\phi_{0,m} = 0$ and use Eq. (3) to find $\bar{n}_{k,m}$. Considering the fact that $\bar{n}_{s,m}$ is exponentially distributed over different optical modes, the PDF for the phase measurement accuracy can be calculated by integrating the PDF given in Eq. (8) over $\bar{n}_s$ weighted with an exponential PDF. Finally, we integrate over all possible measured amplitudes $A$

$$P_\phi(\phi) = \iint \frac{e^{-\bar{n}_{s,m}/\langle \bar{n}_{s,m} \rangle}}{\langle \bar{n}_{s,m} \rangle} A_m P_\Delta \left( A_m \cos\phi_m \left| \frac{\bar{n}_{r,m}}{4} + \frac{\bar{n}_{s,m}}{4} + \frac{\sqrt{\bar{n}_{r,m}\bar{n}_{s,m}}}{2}, \frac{\bar{n}_{r,m}}{4} + \frac{\bar{n}_{s,m}}{4} - \frac{\sqrt{\bar{n}_{r,m}\bar{n}_{s,m}}}{2} \right. \right) $$
$$\times P_\Delta \left( A_m \cos\phi_m \left| \frac{\bar{n}_{r,m}}{4} + \frac{\bar{n}_{s,m}}{4}, \frac{\bar{n}_{r,m}}{4} + \frac{\bar{n}_{s,m}}{4} \right. \right) dA_m d\bar{n}_{s,m}$$

(10)

where $\langle \bar{n}_{s,m} \rangle (= \bar{n}_s/M)$ denotes the average signal strength in a single optical mode.

Figure 3 presents the excellent agreement between the experimental phase error distribution $P_\phi(\Delta\phi)$ and the theoretical distribution (Eq. (10), evaluated numerically). The experimental phase error distribution is obtained by comparing the phase map measured at the low photon limit with the reference phase map measured at $\bar{n}_s/M \sim 2000$ where the impact of shot noise is negligible. As expected, the phase measurement accuracy is significantly reduced as the number of signal photons decreases below the level of $\bar{n}_s/M \sim 1$. However, the phase error is favorably populated around 0 even down to $\bar{n}_s/M \sim 0.004$. The results evidence that even though each DOPC pixel harvests wavefront information from less than a quantum of signal photon energy on average, the information is still useful in steering the output phasor, such that in aggregate we can achieve a significant contrast in time-reversed light focusing through a scattering medium.

For a given number of signal photons, the contrast that can be achieved with optical phase conjugation is much higher than what can be achieved with feedback-based methods. For feedback-based wavefront shaping, $M$ different interferometric measurements are needed, and the maximum enhancement is proportional to the average number of photons recorded per measurement, i.e. $\eta_{max} = \bar{n}_s/M$, with $\bar{n}_s$ the total photon budget for all measurements [25]. Note, however, that feedback-based wavefront shaping allows one to simultaneously determine the wavefronts for focusing at any desired point [26], whereas optical phase conjugation only finds the solution for a single focus at the time. Therefore, for measuring the full transmission matrix of a sample (with all possible combinations of $M$ input modes and $M$ output modes), both methods are expected to reach the same fidelity for a given total photon budget.

To summarize, our study provides the first comprehensive understanding of the interplay between the number of controllable modes and photon budgets available in digital optical phase conjugation. As indicated in the Eq. (1), in general, having one of the three parameters ($M$, $\bar{n}_r$, and $\bar{n}_s$) much larger than the other does not significantly improve the enhancement $\eta$. In other words, the enhancement $\eta$ is capped to the least of three parameters.

Our results, which are directly applicable to the conventional analog phase conjugation scheme [15–18], exhibit numerous counterintuitive implications:

1. It is always beneficial to perform phase conjugation with a resolution as high as possible, even if doing so decreases the number of photons per controlled mode to far below 1. Our results guarantee that this configuration provides an enhancement equal to the total number of photons $\bar{n}_s$ collected during the phase conjugation process.

2. In the low photon limit, the enhancement, traditionally known to solely depend on the number of controlled optical modes, becomes proportional to the parameters affecting $\bar{n}_s$, such as the incident power of the input beam, the exposure time for wavefront measurement, and the tagging efficiency of the guide star (e.g. diffraction efficiency of the ultrasound guide star [3–5] and quantum yield of fluorescence guide star [6]).

3. The number of target modes (i.e. the number of optical modes in guide star(s)) can be increased without compromising the enhancement. In the low photon limit, the effect of the increased photon number $\bar{n}_s$, that results from increasing the number of target modes, will counteract the well-known inverse scaling of the enhancement with the number of target modes [19].

Our results demonstrate that phase conjugation can be used even when the photon budget is extremely low, such as in high-speed focusing through dynamic media, or imaging deep inside tissue. We anticipate the analysis and implications, presented in this study, will serve as a stepping stone towards the adoption and improvement of optical phase conjugation in realizing its biomedical applications.

# Acknowledgements

We thank Haowen Ruan for help on the experiment. This work is supported by the National Institutes of Health (1DP2OD007307-01), the National Institutes of Health BRAIN Initiative (1U01NS090577-01), and a GIST-Caltech Collaborative Research Proposal (CG2012). IMV is supported by the European Research Council under the European Union's Horizon 2020 Programme / ERC Grant Agreement no [678919].

# Footnotes

†Present address: Department of Physics, Korea University, Seoul 136-701, Korea

* chyang@caltech.edu

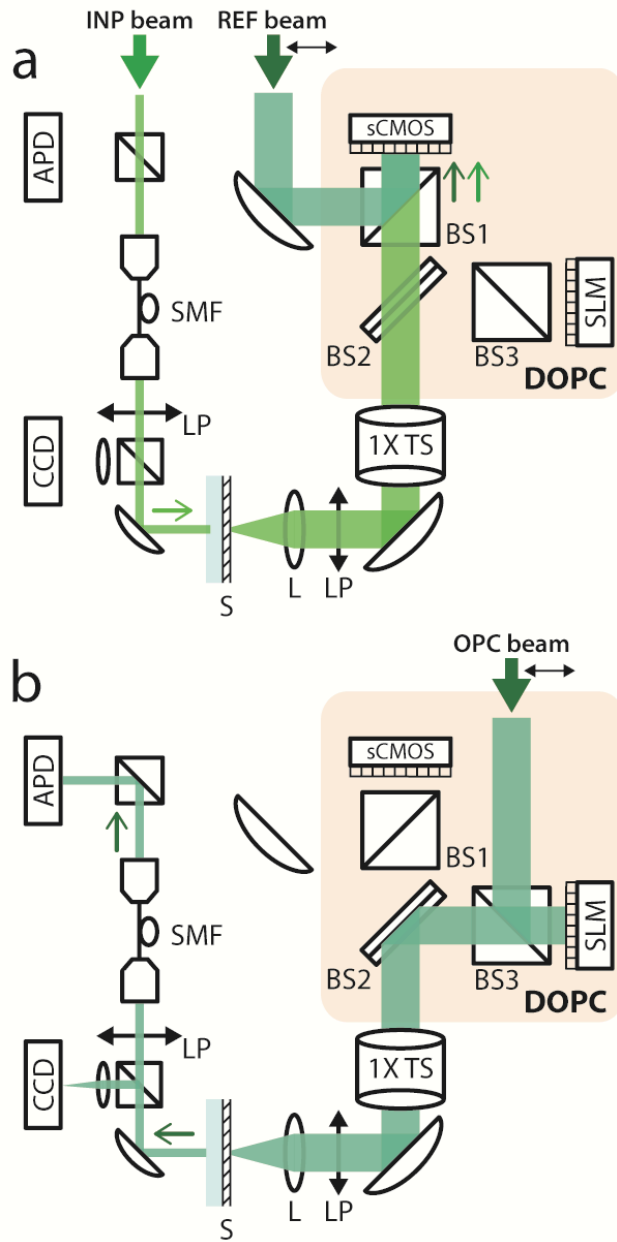

**FIG. 1. Experimental setup.** (a) In the recording step, the input beam (denoted as 'INP beam'), which is spatially filtered through a single-mode fiber (SMF), propagates through the scattering media (S). The scattered input beam is then collected by the plano-convex lens (L, focal length = 5 cm) and relayed to the sCMOS sensor through a 1X telescope system (1X TS), interfering with the collimated reference beam (denoted as 'REF beam'). Complex field distribution of the signal beam is then measured using 4-step phase stepping method. (b) In the playback step, the collimated playback beam (denoted as 'OPC beam')

reflects off the spatial light modulator (SLM) on which the conjugated phase of the measured wavefront is displayed. APD measures the fraction of power in the phase-conjugate beam that is coupled back to the single mode fiber (SMF) and CCD camera directly captures the transmitted intensity distribution. BS, 50/50 beam splitter; LP, linear polarizer.

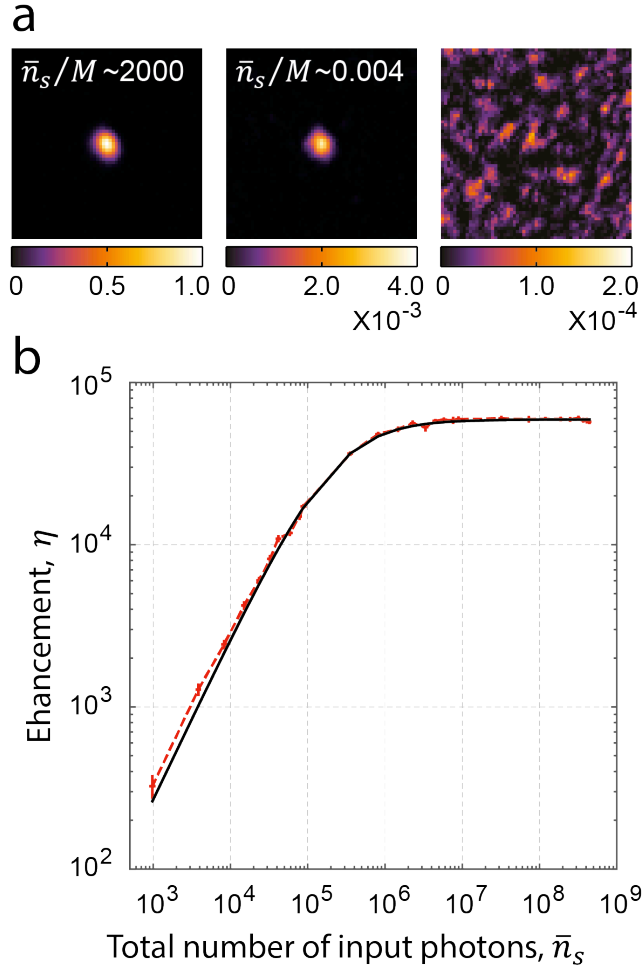

**FIG. 2. Effect of shot noise on time-reversal fidelity.** (a) The intensity distribution of the phase conjugate beam was directly visualized by the CCD sensor. Left and middle: with the phase conjugate beams generated at $\bar{n}_s/M$ of ~2000 and ~0.004. Right: with an unshaped incident beam (i.e. plane pattern on SLM). (b) Experimental and theoretical enhancement versus the total number of input photons $\bar{n}_s$. The solid curve and dotted curve respectively represent the theoretical and experimental enhancement.

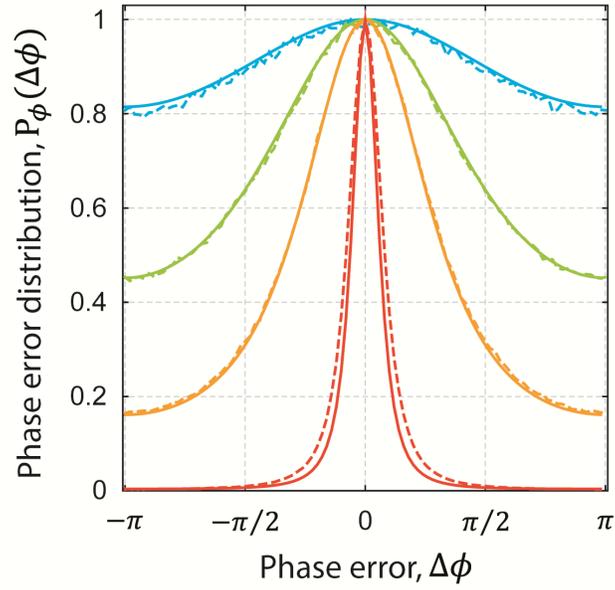

**FIG. 3. Effect of shot noise on phase measurement accuracy.** The phase error distribution, $P_\phi(\Delta\phi)$, was obtained by comparing the phase map measured at a low photon limit (Red, $\bar{n}_s/M \sim 20$; Orange, $\bar{n}_s/M \sim 0.4$; Green, $\bar{n}_s/M \sim 0.07$; Cyan, $\bar{n}_s/M \sim 0.004$) with the reference phase map measured at a negligible shot noise level ($\bar{n}_s/M \sim 2000$). The solid curve and dotted curve respectively represent the theoretical and experimental phase error distribution. All curves are normalized such that the maximum value is unity.